\documentstyle[psfig]{europhys}

\def\And{{\rm and\ }}

\def\stars{\bigskip\centerline{***}\medskip}

\newif\ifboo \boofalse

\begin{document}
\euro{29}{2}{123}{1995}
\Date{??}
\shorttitle{T.~P\"oschel and J.~A.~Freund. FINITE SAMPLE DISTRIBUTIONS}

\title{Size segregation and convection}

\author{T. P\"oschel\inst{1} \And H.J.~Herrmann\inst{2}}
                                                                               
\institute{ 
\inst{1}H\"ochstleistungsrechenzentrum, KFA, D--52425 J\"ulich,
Germany \\
\inst{2}P.M.M.H., URA CNRS 857, E.S.P.C.I., 10 rue Vauquelin, 75231 Paris
Cedex 05, France  
}


\pacs{
\Pacs{46}{10}{Mechanics of discrete systems} 
\Pacs{91}{45}{Convection currents}
\Pacs{64}{75.+g}{Solubility, segregation, and mixing }
}

\maketitle
\begin{abstract}
The size segregation of granular materials in a vibrating container is
investigated using Molecular Dynamics. We find that the rising of
larger particles is accompanied by the existence of convection cells
even in the case of the lowest possible frequencies. The convection
can, however, also be triggered by the larger particle itself. The
possibility of rising through this mechanism strongly depends on the
depth of the larger particle.
\end{abstract}

One of the most puzzling phenomena encountered in granular matter is
size segregation: When a mixture of grains of the same material (equal
density) but different size is shaken in a container the larger
particles rise to the top. This effect has been extensively studied
experimentally \cite{1,2,3} and has much importance in numerous
industrial processes\cite{4}.  Recently this so called ``Brazil nut
effect'' has attracted much interest among physicists\cite{5}.
 
Size segregation inevitably seems to contradict equilibrium
statistical mechanics since the density of the overall packing
increases with polydispersivity and so gravity
makes situations with larger particles on
the bottom energetically more favourable.  Rosato et
al.\cite{6} proposed a Monte Carlo algorithm and put forward a kinetic
argument to explain segregation using the fact that smaller particles
are more mobile. In the same year Haff and Werner\cite{7} did
Molecular Dynamics simulations of rather small systems and claimed
that segregation was essentially a consequence of solid friction and
the rotation of the particles.  Jullien et al.\cite{8} used a piling
technique which is non--stochastic as compared to the one of ref.~6 and
found a critical ratio $\cal R$ for the radia of spherical particles below
which no segregation occurs.  Based on these ideas Duran et
al.\cite{9} formulated a geometrical theory for segregation in which
the small particles glide down along the surface of the larger
particle, the critical ratio $\cal R$ of radia being between continuous and
discontinuous gliding. They also claimed experimental evidence for two
types of dynamics and visualized the discontinuous ascent of the
larger particle through stroboscopic photos. Jullien et al.\cite{10}
reproduced the discontinuous dynamics by including horizontal random
fluctuations into their model.

Parallel to these local theories there has been the ``convection
connection'': It is known experimentally\cite{11} and
numerically\cite{12,13} that shaken assemblies of spheres form
convection rolls which are attached to the walls of the container. For
weak shaking the convection rolls only appear on the surface.  Knight
et al.\cite{14} showed experimentally that segregation was due to this
convection and the fact that larger particles have a harder time
entering again into the bulk once they are on the surface. They also
verified an exponential decay of the convection strength as function of depth
for weak shaking\cite{15}. Duran et al.\cite{16} verified segregation
due to convection in two dimensions for strong shaking and claimed
that above mentioned local mechanisms are at work at weak
shaking. Their stroboscopic pictures showed convection cells
above the particles.

Here we present large scale Molecular Dynamics simulations showing
that also for weak shaking convection is responsible for segregation
but in a more intricate way: Under certain conditions
the larger particle is able to pull down the convection rolls 
due to the more efficient momentum transfer 
and then rises within its flow.  Because
of the exponential decay of convection with depth the ability to rise
critically depends on the vertical position of the larger particles.

We used the classical Gear predictor--corrector Molecular
Dynamics\cite{17} algorithm and the contact forces proposed by
Cundall and Strack\cite{18}: Two
grains $i$ and $j$ at positions $\vec{r}_i$ and $\vec{r}_j$ interact if
the distance between the center points is smaller than the sum of their radii
$\mid \vec{r}_i-\vec{r}_j \mid~<~R_i + R_j$.
For this case the force between particles $i$ and $j$ moving with the velocities
$\dot{\vec{r}}_i$ and $\dot{\vec{r}}_j$ and rotating with angular velocities
$\dot{\Omega}_i$ and $\dot{\Omega}_j$ is given by 
\begin{equation}
\vec{F}_{ij} =  F_{ij}^N \cdot \frac{\vec{r}_i
-\vec{r}_j}{|\vec{r}_i-\vec{r}_j|} + F_{ij}^S \cdot  \left({0 \atop 1} ~{-1
\atop 0} \right) \cdot \frac{\vec{r}_i - \vec{r}_j}{|\vec{r_i}-\vec{r}_j|}
~. 
\end{equation}
with the normal and shear forces being
\begin{eqnarray}
F_{ij}^N  &=& Y \cdot \left(R_i + R_j - |\vec{r}_i - \vec{r}_j|\right)+
        ~\gamma_N \cdot m_{ij}^{eff}\cdot |\dot{\vec{r}}_i - \dot{\vec{r}}_j|\\
F_{ij}^S &=& sign\left(\left| \vec{v}_{ij}^{\,rel}\right| \right)
\min \left\{\gamma_S \cdot m_{ij}^{eff} \cdot 
|\vec{v}_{ij}^{\,rel}|~,~ \mu \cdot  
        |F_{ij}^N| \right\} ~.
\label{eq_coulomb}      
\end{eqnarray}

The relative velocity $\vec{v}_{ij}^{\,rel}$ between $i$ and $j$
and the effective mass $m_{ij}^{eff}$ of the
particles $i$ and $j$ are defined as
\begin{eqnarray}
\vec{v}_{ij}^{\,rel} &=& (\dot{\vec{r}}_i - \dot{\vec{r}}_j) + R_i \cdot
\dot{\Omega}_i + R_j \cdot \dot{\Omega}_j \\
m_{ij}^{eff} &=& \frac{m_i \cdot m_j}{m_i + m_j} ~.
\label{eq_eff_mass}
\end{eqnarray}
The resulting momenta $M_i$ and $M_j$ acting upon the  particles are
\begin{equation}
M_i = F_{ij}^S \cdot R_i~~,~~
M_j = - F_{ij}^S \cdot R_j ~.
\end{equation}
Eq.~(\ref{eq_coulomb}) takes into account
that the particles slide on each other for the case that the
inequality $\mu \cdot \mid F_{ij}^N \mid~<~\mid F_{ij}^S \mid$ holds,
otherwise they feel some viscous friction.

Throughout our simulation we used the parameters
$Y=3\cdot10^6~g/sec^2$ (Young modulus), $\gamma_N=100~sec^{-1}$,
$\gamma_S=1~sec^{-1}$ (phenomenological normal and shear friction
coefficients) and $\mu=0.5$ (Coulomb friction parameter). We considered
$N=950$ particles with radii uniformly distributed in the
interval $R_i \in \left[0.85,1.15 \right]~cm$  and with 
masses $m_i=2 \pi R_i^2 \rho$, $\rho=1~g/cm^2$. The particles are put into a
two--dimensional box having walls made of particles with the
same material characteristics as the grains and which vibrates
vertically according to $y_{box}=A\cdot \sin (2\pi f t)$
with $A=2~cm$. Gravity acts in negative $y$--direction
$F_i^{gr}=-m_i~g$, $g=981 cm/sec^2$. The time step for the numerical
integration of the Gear predictor--corrector scheme of 5th order was
$\Delta t= 5\cdot10^{-5} sec$.

\begin{figure}
\centerline{\psfig{figure=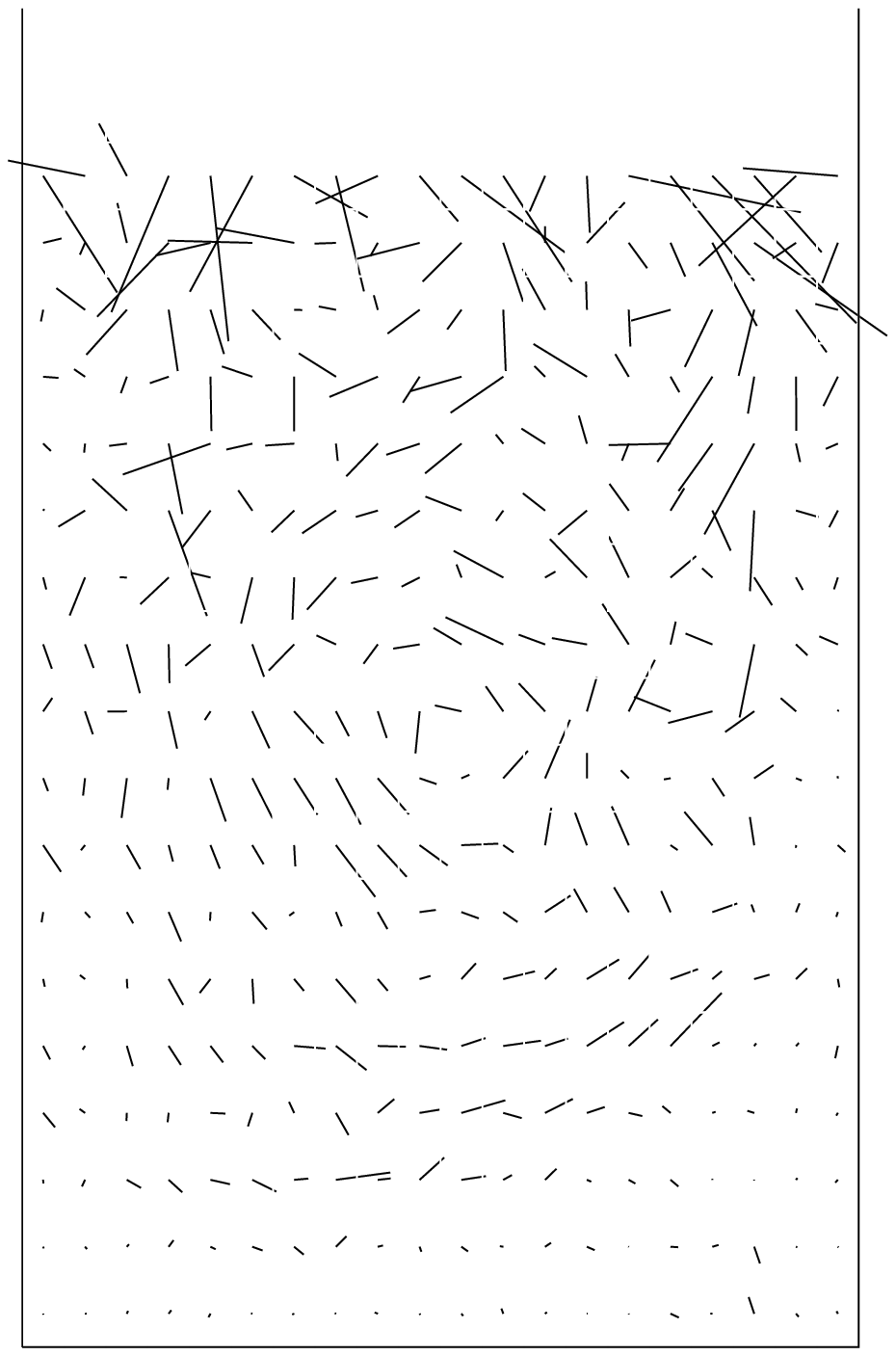,width=3.0cm} \hspace{0.5cm}
           \psfig{figure=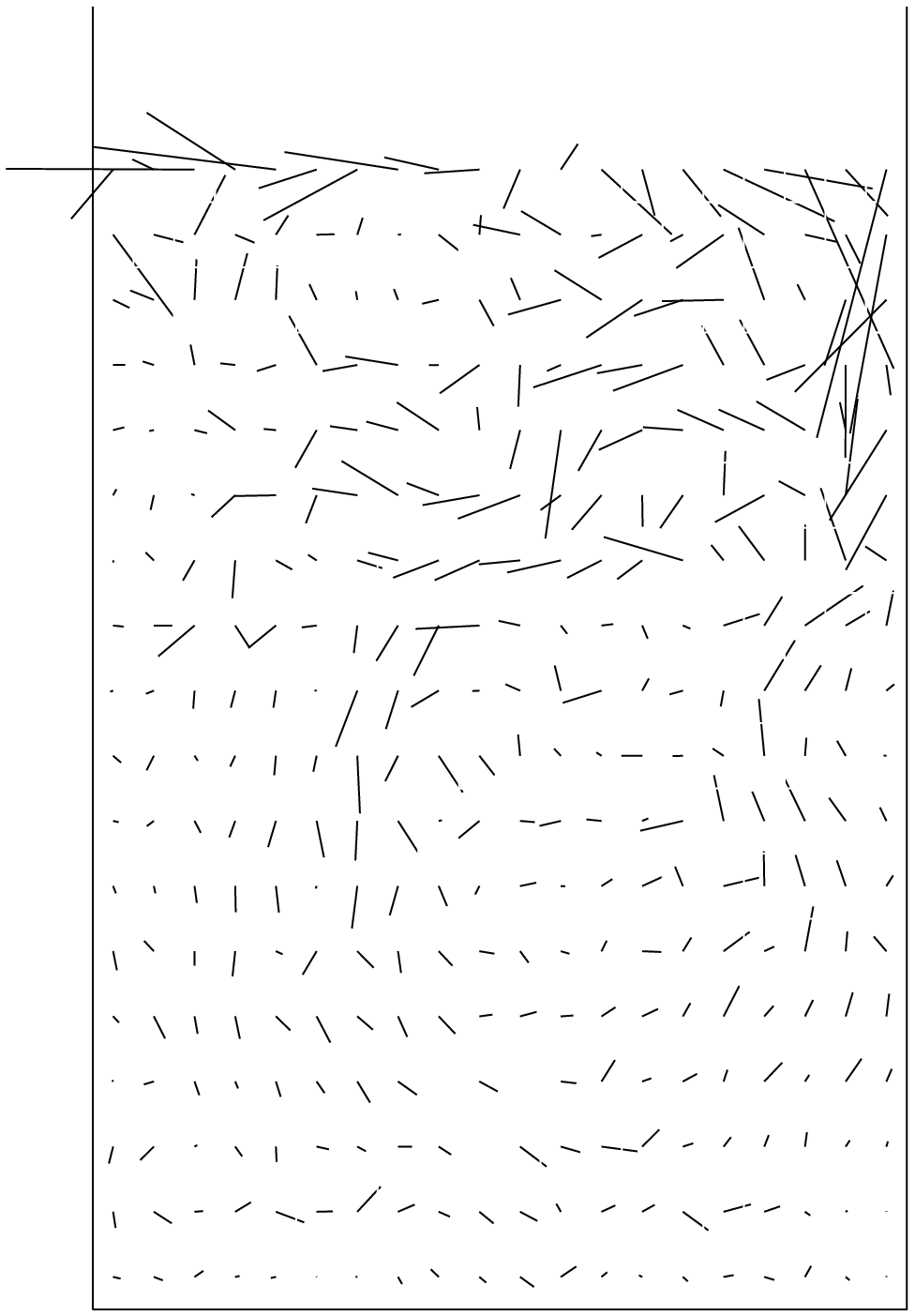,width=3.0cm}} \vspace{0.5cm}
\centerline{\psfig{figure=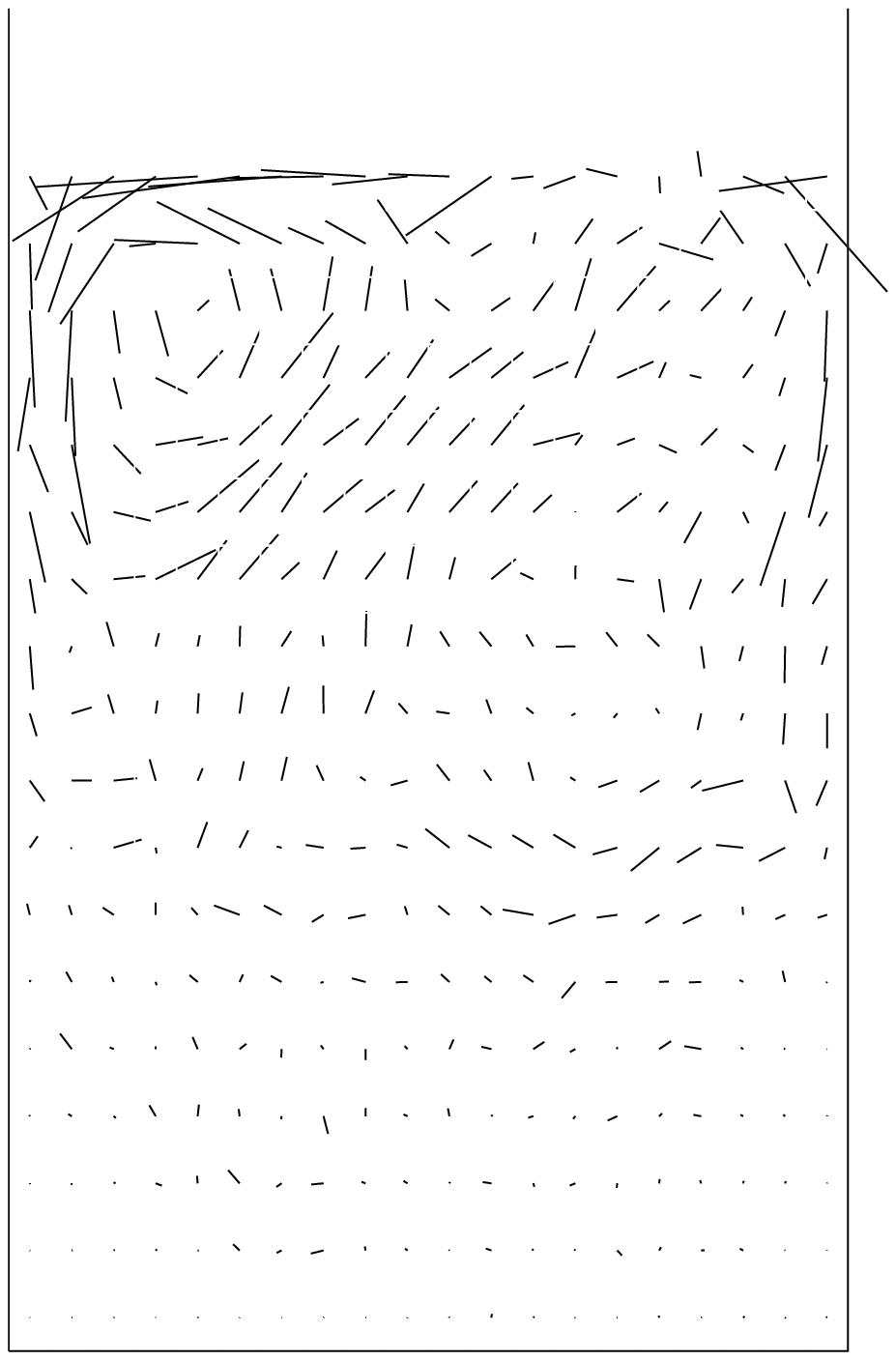,width=3.0cm} \hspace{0.5cm}
           \psfig{figure=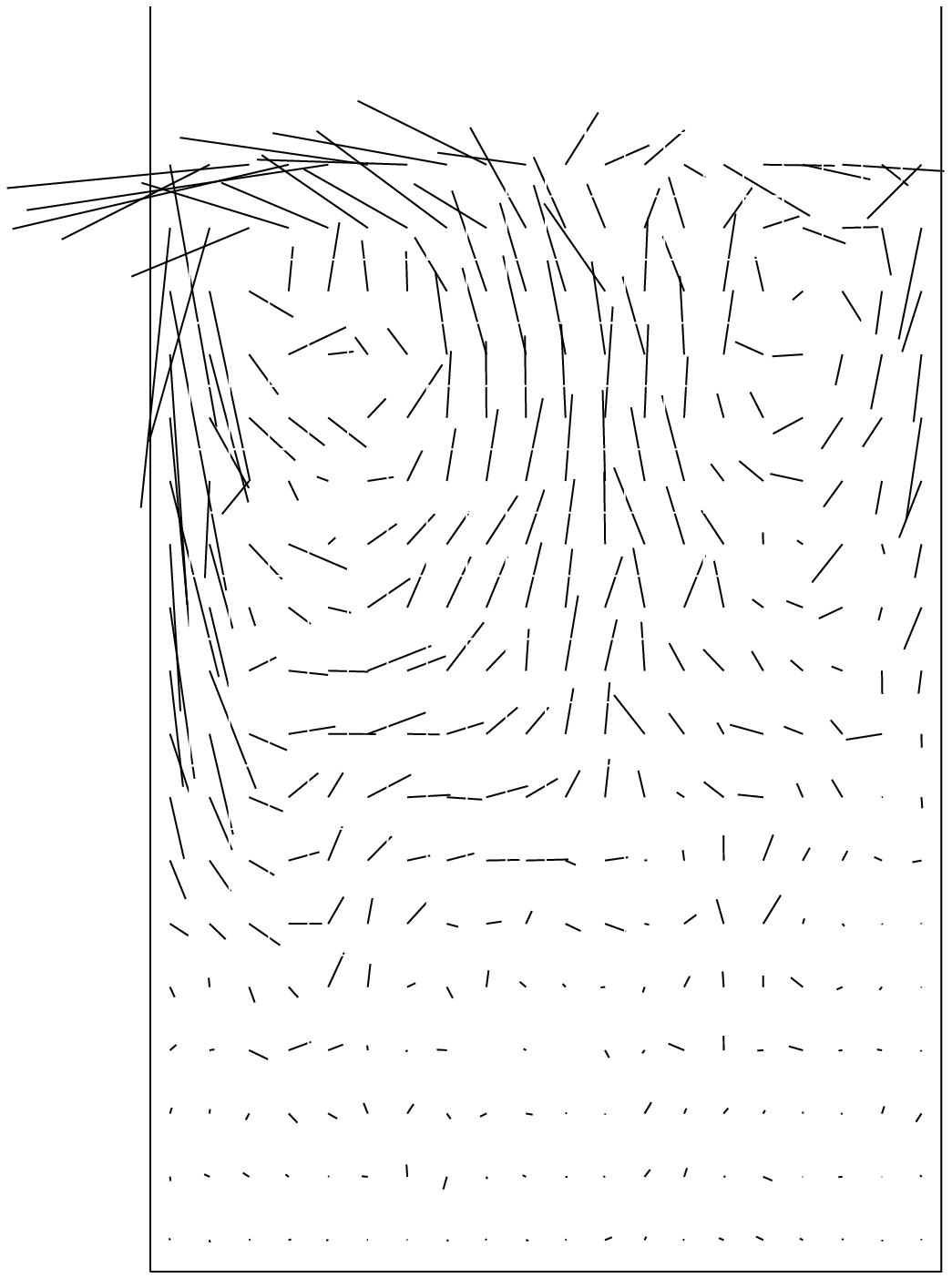,width=3.0cm}}\vspace{0.5cm}
\centerline{\psfig{figure=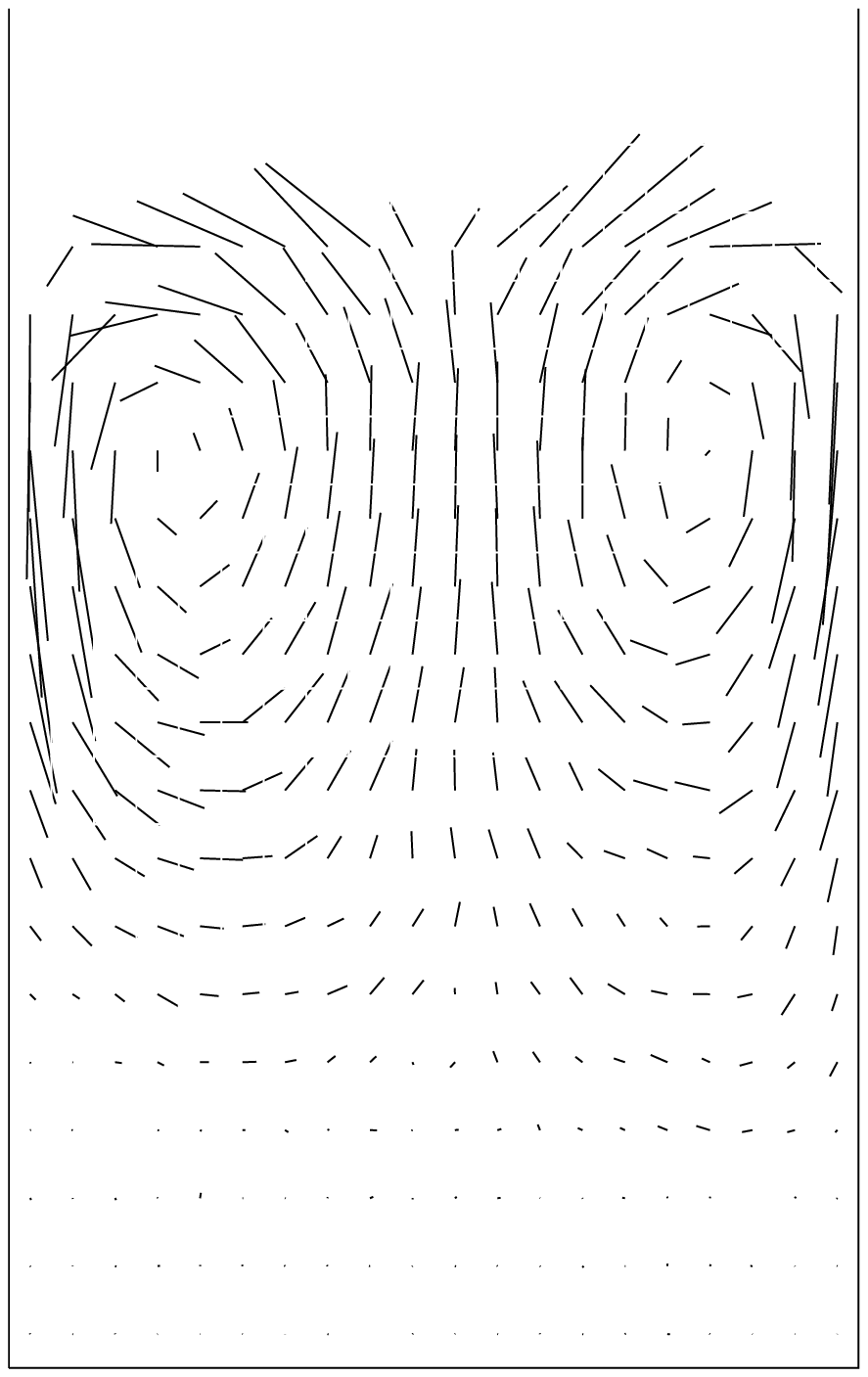,width=3.0cm} \hspace{0.5cm}
           \psfig{figure=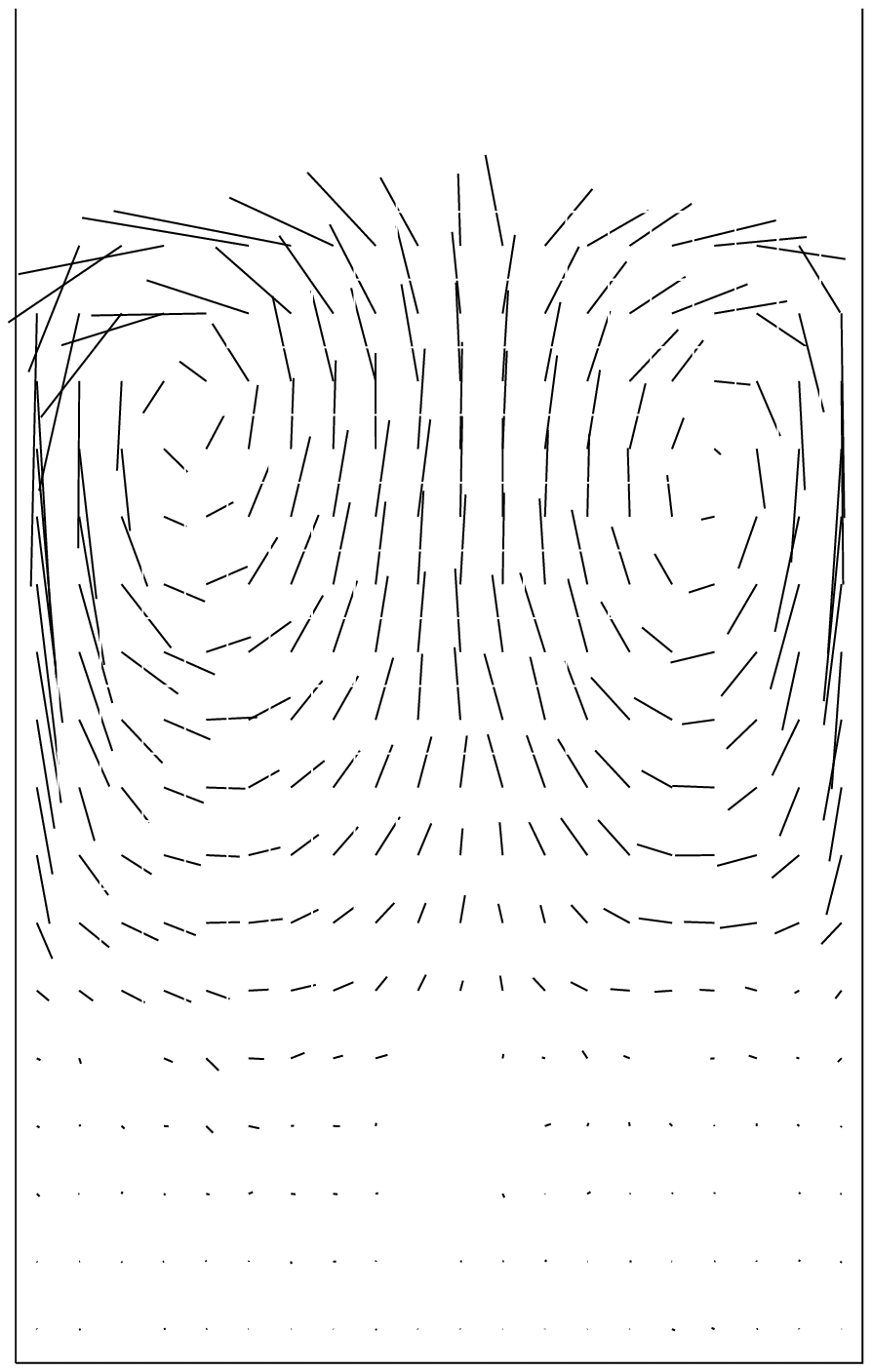,width=3.0cm}\vspace{0.3cm}}
\caption{Convection rolls in systems without (left) and with the
larger particle (right) for three different shaking frequencies,
$f=2.6~sec^{-1}$ (upper figure), $f=2.8~sec^{-1}$ (central figure) 
and $f=3~sec^{-1}$ (lower figure).
In the central case the big particle triggers convection
rolls. The pictures were obtained by averaging over 50
shaking periods of the box.}
\label{1}
\end{figure}

We investigate segregation and convection behaviour as function of
the vibration frequency $f$ in two different systems, 
either all particles are small ($R_i \in \left[0.85,1.15 \right]~cm$) 
or we add one single big particle of $R_1=4~cm$
located in the centre of the box close to the bottom. 
In order to investigate closer what happens at the onset of segregation
we keep all the other parameters fixed.
Fig. \ref{1} shows the convection cells without the larger particle
(left) and with the larger particle (right) for three different
frequencies $f=2.6~sec^{-1}$ (upper figure), $f=2.8~sec^{-1}$ 
(central figure) and $f=3~sec^{-1}$ (lower figure). 
For $f=2.8~sec^{-1}$ the convection cells of
both systems, with and without the big particle, differ significantly
while they are quite similar in the other cases. This indicates that
there is a certain frequency interval where the presence of the big
sphere triggers the onset of convection which finally leads to
segregation. Indeed we find that convection is always present when
segregation happens. It is important to note that if one is close enough
to the onset of segregation just putting
the larger particles one row lower can entirely supress the
effect of segregation. This dependence of
segregation on the height is quite strong and has not been discussed
in the literature. By changing the frequency $f$ very slowly and
measuring the convection flow through a plane at a certain height we
observed that the transition from the fluctuation regime to the
convection regime is very sharp and differs between different runs
for less than $\Delta f=0.05~sec^{-1}$. Moreover when increasing the
frequency the transition occurs almost exactly at the same frequency
as when decreasing the frequency, i.e. there is no hysteresis. This is
in contrast to the case of viscous fluids which tuned
out of equilibrium by applying a temperature gradient
have nonequilibrium dissipative
structures, e.g. B\'enard convection, with a characteristic
hysteresis.

\begin{figure}
\centerline{\psfig{figure=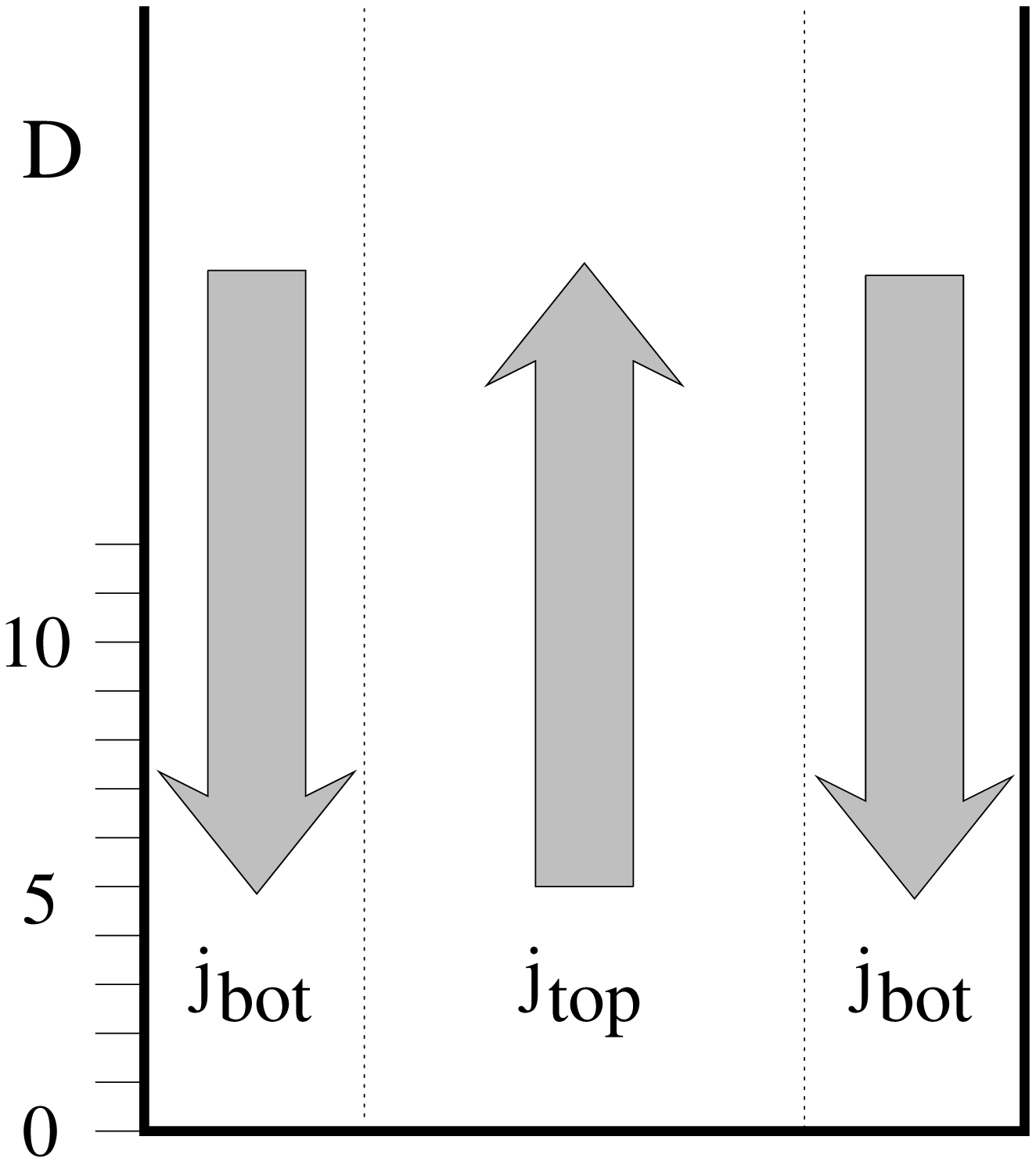,width=3.5cm}\vspace{0.3cm}}
\caption{Schematic diagramm illustrating how the material 
flow $j_{bot}$ close by the walls and $j_{top}$ in the center
were added to obtain the convective flow $\Phi$. }
\label{2}
\end{figure}

\begin{figure}
\centerline{\psfig{figure=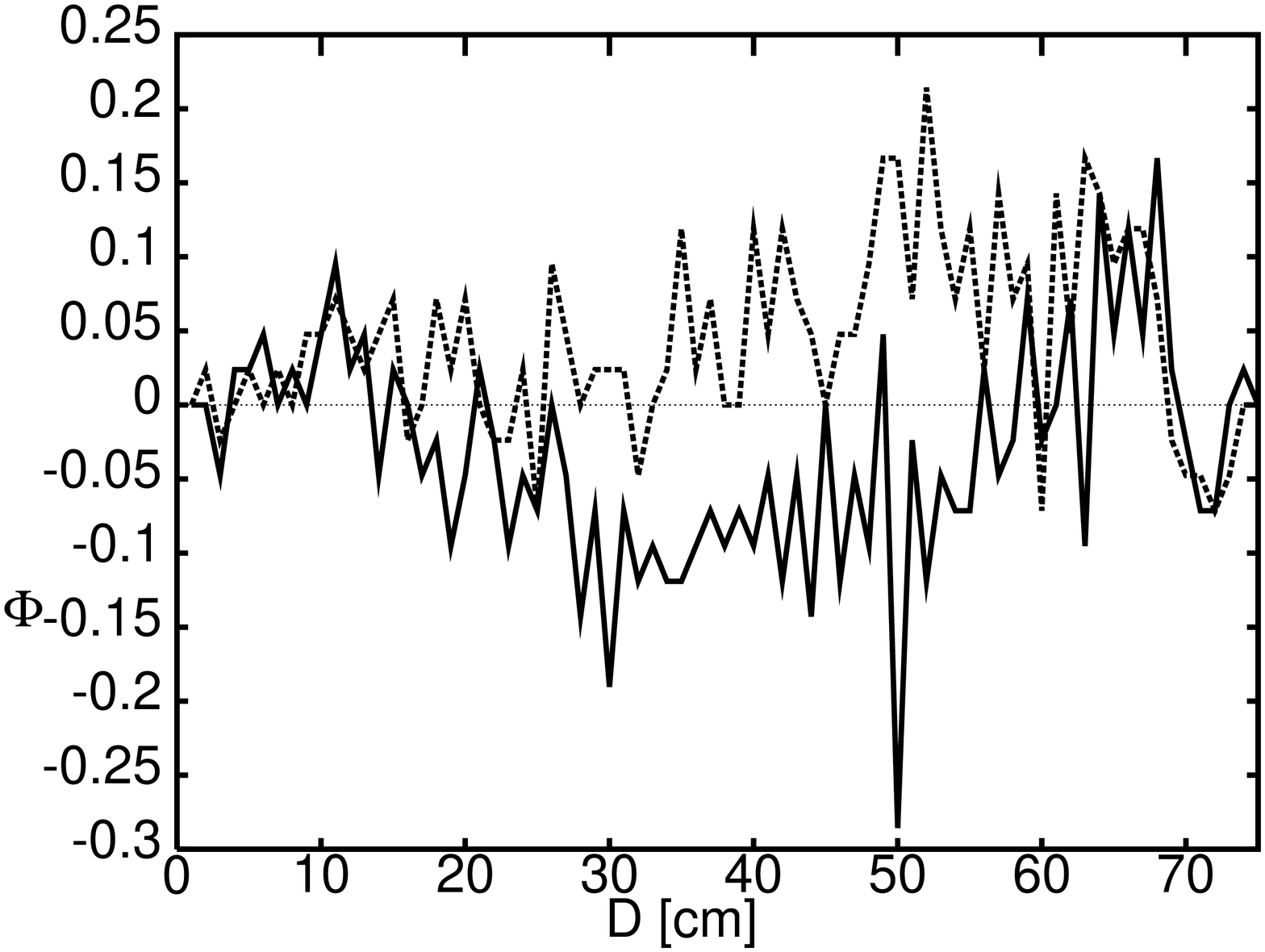,width=6cm,angle=270}}\vspace{0.5cm}
\centerline{\psfig{figure=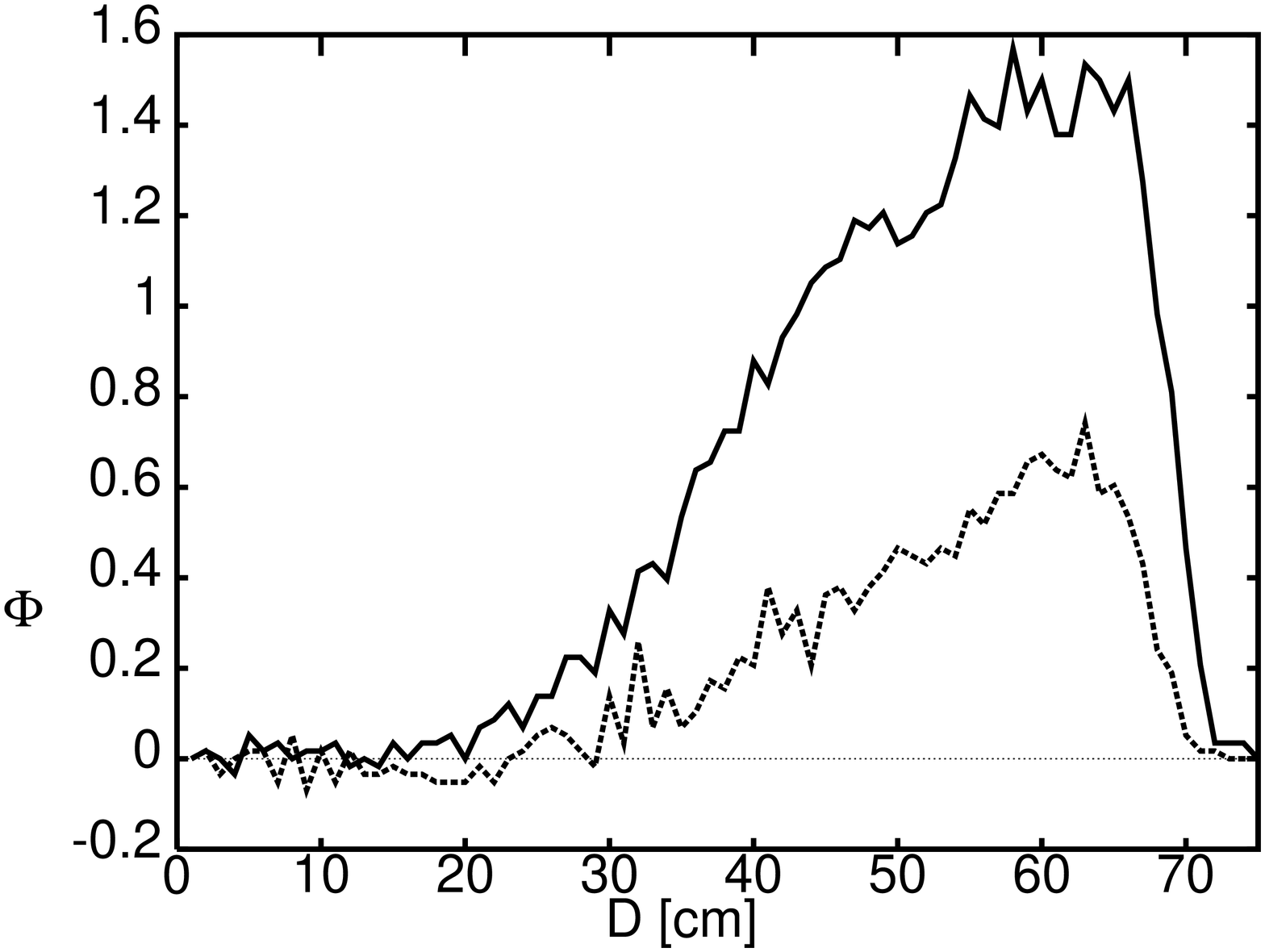,width=6cm,angle=270}}\vspace{0.5cm}
\centerline{\psfig{figure=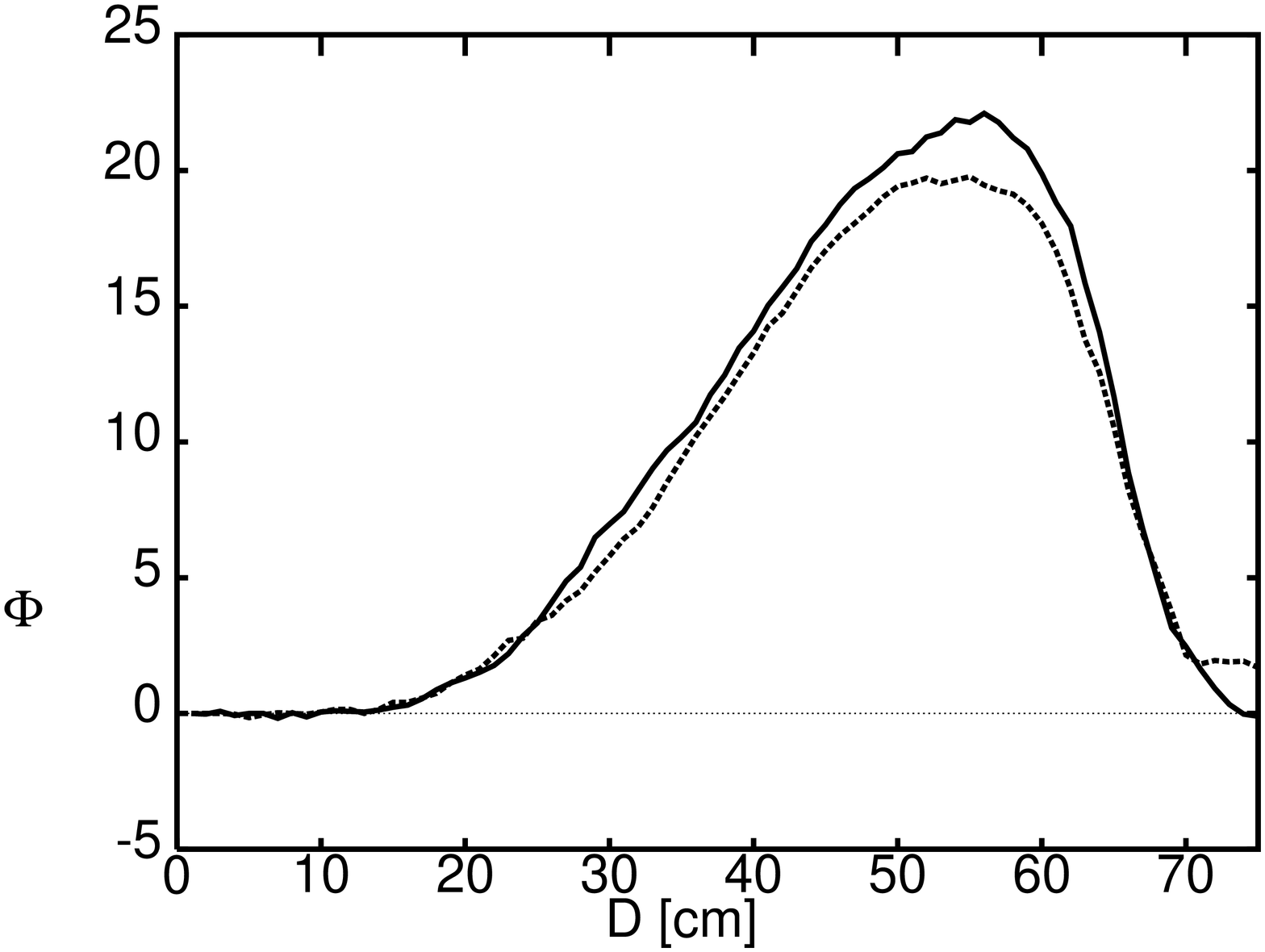,width=6cm,angle=270}\vspace{0.3cm}}
\caption{Strength of the convection rolls measured through the 
flux $\Phi$ as a function of the
height $D$, $f=2.6~sec^{-1}$ (upper figure), $f=2.8~sec^{-1}$ 
(central figure) and $f=3~sec^{-1}$ (lower figure). The height $D$ is 
measured such that the bottom is at the origin of the axis.
The flux $\Phi$ is measured in units of particles per period.
In the central case the convection rolls are stronger and reach
deeper inside the material in the presence of the big particle.
This triggered convection roll catches the big particle
and forces it to rise to the top. }
\label{3}
\end{figure}

The triggering of convection cells by the big particle is investigated more
quantitatively by calculating the convective flux $\Phi$ defined as
the sum of material (mass) flow in the centre of the box 
$j_{top}$ and the flow close to the walls $j_{bot}$
by considering that these flows have opposite signs as illustrated
in Fig. \ref{2}. The flows $j_{top}$ and $j_{bot}$ are defined by
adding the number of particles which move in one direction minus
the ones moving in the opposite direction and is measured in particles
per cycle time of the box. In fact we measure for each particle if
the positions at subsequent nodes of the vibration are on different 
sides of a height line, where
the height of the box was divided into 80 height lines between the
bottom of the box and the surface of the packing.
Fig. \ref{3} shows the convective flux $\Phi$
through planes at different heights $D$ for both systems and for the three
different frequencies. For $f=2.6~sec^{-1}$ we find almost no directed
flow but only fluctuations. For $f=3~sec^{-1}$ both systems behave similarly,
as could also be observed in Fig. \ref{1}. For $f=2.8~sec^{-1}$,
however, the convection cells clearly
extend deeper due to the existence of the larger particle. Apparently
the larger particle is able to pull the convection cells down.

This effect can be explained by the fact that in the region
around the large particle the accelerations are higher since the
momentum is transferred with less dissipative loss through the larger
particle than through a corresponding pile of smaller particles of
the same volume. We measured the sum of the absolute values of the 
forces of all the particles in a region around the position of the
large particle and averaged it over time. The region was ring shaped
with the inner border of radius $4.5~cm$ and the outer border of radius
$8~cm$. At the onset of segregation ($f = 2.8~sec^{-1}$)
the average force of the small particles around the large particle 
is about 15\% larger than that of the small 
particles in the same region in a system containing no large particle.
This effect is strongest in the lower part of the ring shaped region.
For $f = 2.6~sec^{-1}$ the difference is only 5\%.
Therefore the accelerations in the region
around a large particle are larger than if no particle would be present.
We believe that this increase in high frequency oscillations
is responsible for pulling the convection rolls down. It is,
however, interesting to note that the granular temperatures in this
region is roughly the same in the two systems.

One can see from Fig. \ref{1} that the convection cells decay very
sharply in strength but that even in the deep regions some essentially
horizontal motion occurs. This is reminiscent of the stroboscopic
pictures of \cite{16} implying that even in the low acceleration regime
some particles move inward horizontally. Within our framework this
motion could be interpreted as the exponentially weak tail of the
convection rolls.

\begin{figure}
\centerline{\psfig{figure=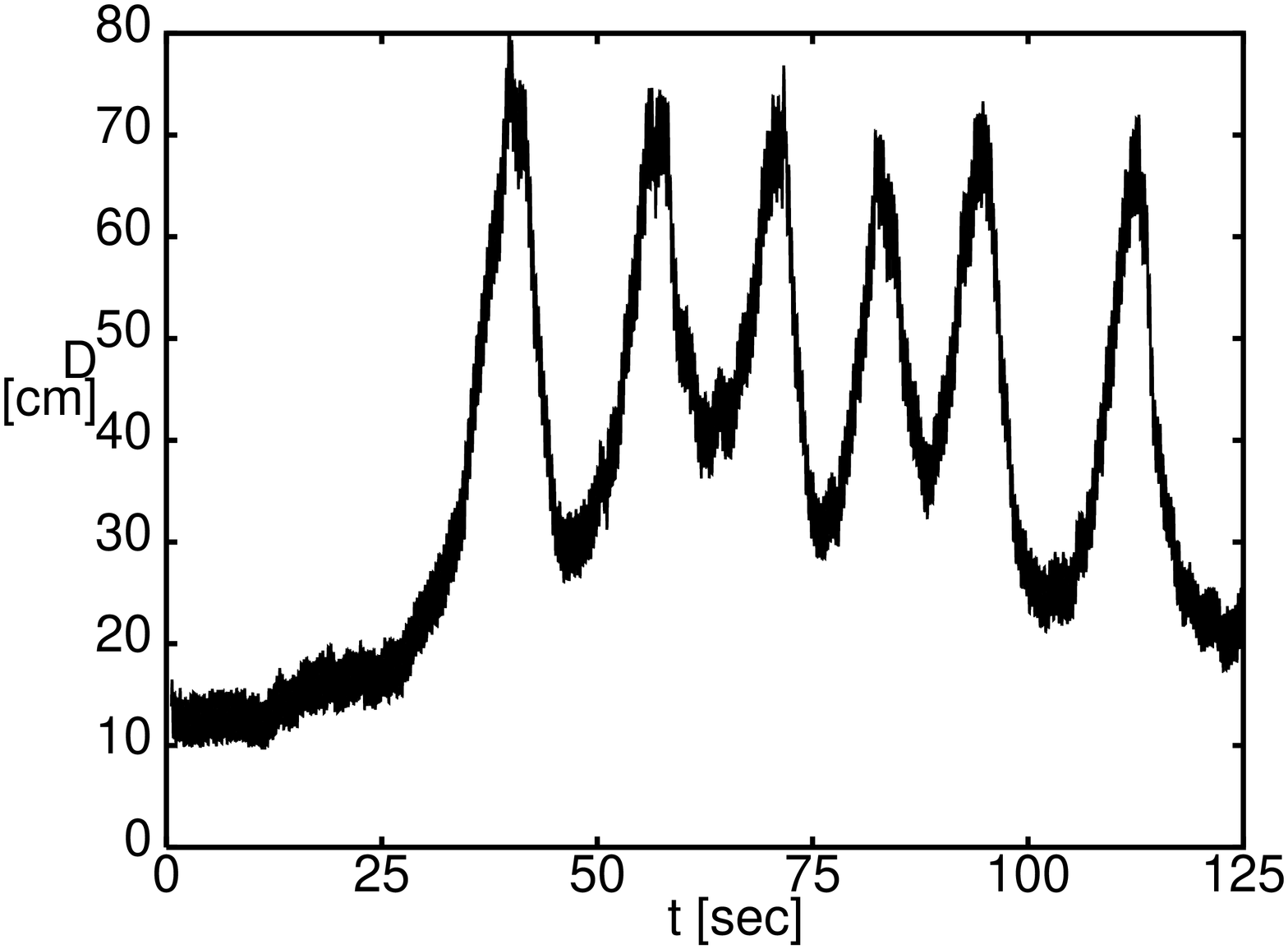,width=6cm,angle=270}
\psfig{figure=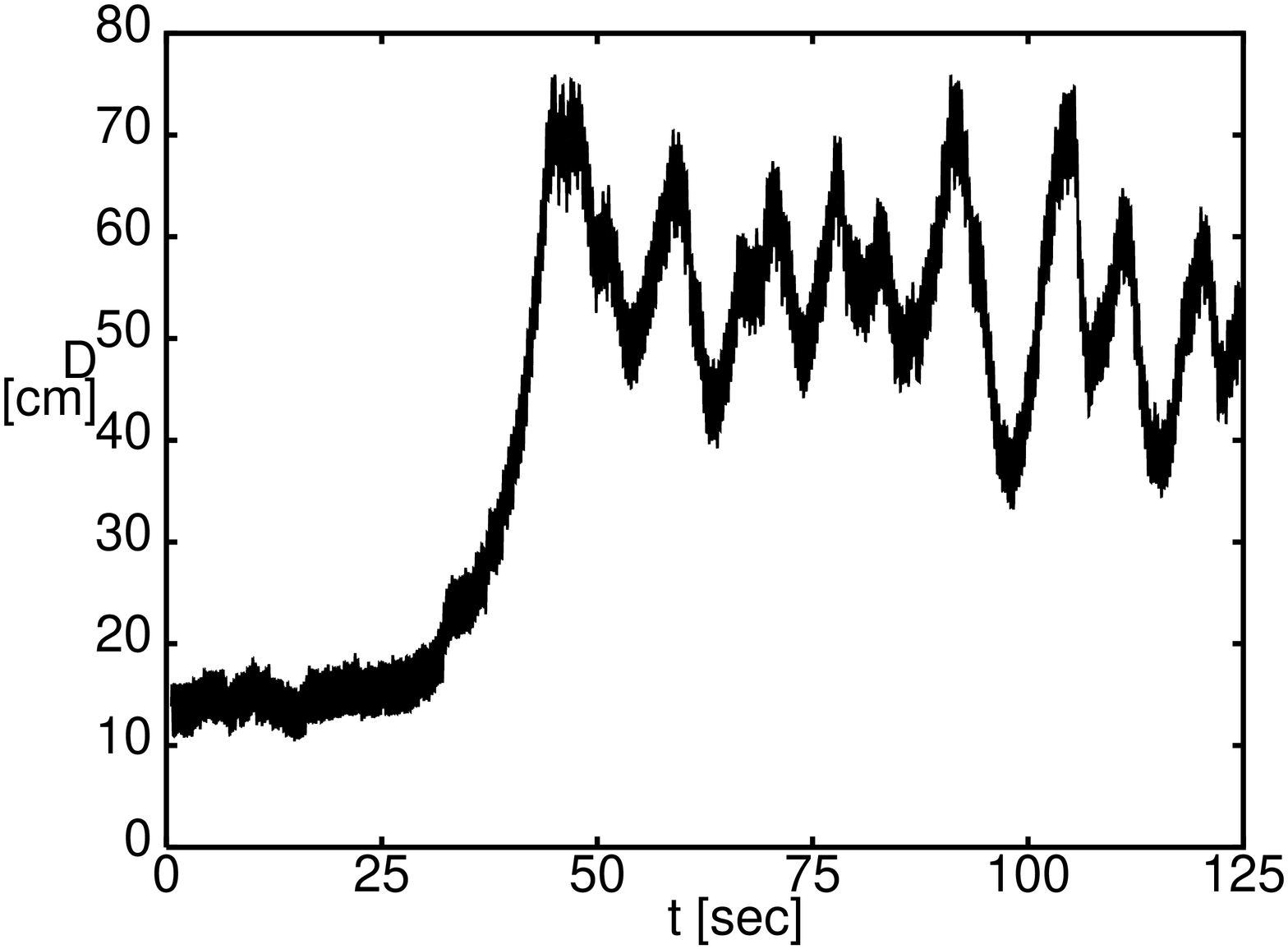,width=6cm,angle=270}\vspace{1cm}}
\caption{Evolution of the vertical position of the big particle
as function of time, (a) ${\cal R} = 2.0$ and (b) ${\cal R} = 3.0$ for
$f=3.2~sec^{-1}$.} 
\label{4}
\end{figure}

Next we investigated the dependence of the onset of convection on
the ratio of radia $\cal R$. Note that in our case ${\cal R} = R_1/1 cm$
because the mean radius of the small particles
is $1 cm$. For $f=3.2~sec^{-1}$ we observed
a big particle of radius ${\cal R} = 4.0$ moving up immediately.
We also studied the cases ${\cal R} = 3.5$, ${\cal R} = 3.0$, 
${\cal R} = 2.5$ and ${\cal R} = 2.0$. 
In these cases the large particle remains a certain waiting
time on the bottom before it suddenly moves up quite rapidly.
Fig. \ref{4} shows a typical evolution
of the vertical position of the big 
particle with radius ${\cal R} = 2.0$ and ${\cal R} = 3.0$. The waiting times
do not noticeably depend on $\cal R$ being of the order of
$30~sec$.
Once the large particle comes to the top 
it performs an oscillating motion
going up and down (whale effect) that has also been observed 
experimentally\cite{20}. This motion seems due to the convection
rolls: In the case where $\cal R$ is smaller the oscillating motion
is more regular because the larger particle has less difficulty
reentering into the bulk from the surface and to follow the convective
motion. For larger $\cal R$ the larger particle has more difficulty
reentering thus leading to a more erratic horizontal motion as seen
in Fig. \ref{4}. Particles with smaller $\cal R$ also seem to dip deeper
into the bulk showing that the convection cells can move them more
efficiently.

We have shown with large numerical calculations of granular media in
a vibrating box that in two dimensions segregation is intimately
connected to convection. The larger particles, surrounded by a
region of higher acceleration, loosen the material thus deepening the
penetration of the convection rolls from the top into the medium.
After some waiting time the 
larger particles are caught by the lower part of the
rolls and pulled up. This triggering effect is only relevant close
to the characteristic onset frequency of segregation which is sharply
defined and strongly dependent on the initial depth of the large
particle. Once the large particle is on the top it periodically
goes up and down (whale effect) driven by the convection cells.

In addition to convection arching and geometry are also important as
pointed out by many authors in the past. How the large particles
move in the exponentially weak convection field before being
lifted upwards and the $\cal R$-dependence of the mobility are
probably best described by Duran et al.'s local arching
mechanisms\cite{9,16}. The fact that the whale effect of larger
particles is less pronounced is certainly due to their lower
mobility because of steric hindrance effects as formulated by
Rosato et al.\cite{6}. 

Many of the details of segregation are still
not completely clear and in particular in three dimensions additional
geometrical effects might play a role. This as well as other questions
are difficult to study conclusively with our numerical technique
due to the excessive requirements in computer time. It would for
instance be interesting to see what happens when the box is so wide that
the walls of the box are much farther away from the large particle
than the height of the packing. In this experimentally relevant case
the walls would not be able to stabilize the convection rolls. 
Simulations with periodic boundary conditions have, however, provided
rather similar results as with fixed boundaries~\cite{19}. It would
also be interesting to study larger ratios $\cal R$ in order to
verify predictions made about a characteristic value of ${\cal R} =
12$~\cite{8,9} but for that case one would need to consider
substantially larger systems.
The limitations in observation
time due to the computational requirements also puts limits
on the determination of the segregation velocity and we cannot
exclude that particles rise
on time scales much larger than the ones accessible numerically.

\stars
H.J.H. thanks Jacques Duran and Daniel Bideau for enlightening
discussions. Stefan Schwarzer and Hans--J\"urgen Tillemans are thanked
for their help.


\end{document}